\documentclass{article}

\usepackage[english]{babel}

\usepackage[letterpaper,top=2cm,bottom=2cm,left=3cm,right=3cm,marginparwidth=1.75cm]{geometry}

\usepackage[sort&compress,super,square,numbers]{natbib}
\usepackage{amsmath}
\usepackage{graphicx}
\usepackage[colorlinks=true, allcolors=blue]{hyperref}
\usepackage{amssymb}
\usepackage[many]{tcolorbox} 
\usepackage{float}
\usepackage{tabularx}
\usepackage{listings}
\usepackage{setspace}
\usepackage{enumitem}
\usepackage[affil-it]{authblk}
\usepackage{natbib}
\usepackage{booktabs}
\usepackage{multirow}
\usepackage{rotating}

\tcbset{
    sharp corners,
    before skip = 0.2cm,    
    after skip = 0.5cm      
}                           
\newtcolorbox{calloutbox}{
    colback=gray!10,
    boxrule = 0pt  
}

\title{Assessing the Impact of Intercurrent Events on Power and Sample Size for Estimands with Time-to-Event Endpoints}

\makeatletter
\renewcommand{\maketitle}{%
  \begin{flushleft}
  \begin{spacing}{2}
    \textbf{\LARGE \@title} \\[2em]
    \@author
  \end{spacing}
  \end{flushleft}
}
\makeatother

\author[1]{Daniel J Bratton}
\author[2]{Fiona Guillard}
\author[1]{Sunita Rehal}
\author[1]{Thomas Drury}

\affil[1]{Quantitative Sciences Innovation, GSK, London, UK}
\affil[2]{Statistics \& Programming, Veramed Ltd., Twickenham, UK}

\date{September 30, 2026}
\begin{document}
\maketitle
\vspace{-2em}  

\begin{calloutbox}
\textbf{Abstract:}
The precise definition of a primary estimand, accounting for intercurrent events (IEs) as per the ICH E9(R1) addendum, is fundamental to the design and interpretation of clinical trials. Conventional power and sample size calculations, however, often do not adequately incorporate the impact of IEs and their corresponding handling strategies, creating a risk of over- or under-powered studies. While simulation-based approaches can address this complexity, they are often computationally intensive and may only explore a limited set of scenarios. In this paper, we introduce a set of formulae for calculating power for estimands with time-to-event endpoints. We focus on estimands that use treatment policy, hypothetical, composite, or a combination of strategies for handling IEs, assuming constant hazards and that a participant's risk of experiencing an IE is unrelated to their risk of other IEs or the primary endpoint, while the effect of an IE on the endpoint is determined by the chosen handling strategy. Simulation is used to assess the accuracy of calculated power in a variety of settings, and we explore deviations in power estimates in scenarios where event times for outcomes and IEs are dependent, hazards are non-constant, or follow-up times vary. We illustrate the practical application of our approach through a case study in nasal polyposis, examining the sensitivity of sample size requirements to varying IE rates and their impacts on post-IE outcomes. The proposed formulae facilitate rapid and accurate power and assurance calculations, enabling clinical trial designs to be more closely aligned with the estimand of interest.
\end{calloutbox}


\section{Introduction}
The estimands framework, introduced in the ICH E9(R1) addendum in 2019, is now well established in clinical trial design and analysis.\cite{ICH_E9R1_2019} By requiring explicit specification of the treatment effect of interest, the framework improves clarity in the formulation, interpretation, and communication of trial objectives. A central component is the structured handling of post-baseline events, termed intercurrent events (IEs), which may affect either the interpretation or the existence of outcome data. 

Before implementation of the addendum, sample size and power were typically determined without explicit consideration of the strategies to be used for handling IEs or assumptions about their frequency and impact. Design choices were often guided by broad analysis concepts, such as intent-to-treat or per-protocol populations, with power calculations informed by historical trials using similar conventions. While pragmatic, these approaches did not always ensure alignment between the design assumptions and the primary scientific question of interest. 

Under the estimands framework, trial design is expected to align with a clearly defined estimand, strengthening coherence between objectives, analysis methods, and design assumptions.\citep{Ratitch2020} The addendum emphasises that ``a precise description of the treatment effects of interest should inform sample size calculations'' and that ``particular care should be taken when making reference to historical studies that might, implicitly or explicitly, have reported estimated treatment effects or variability based on a different estimand''.\citep{ICH_E9R1_2019} This places greater emphasis on understanding how assumptions regarding IEs influence power and sample size requirements. The clinical trial protocol template recently introduced in ICH M11 (Guideline on clinical electronic structured harmonised protocol) also states that details of how IEs are incorporated into the sample size calculations should be included.\citep{ICH_M11_2025} 

Simulation-based methods offer flexibility for evaluating power but can be computationally intensive and may limit extensive sensitivity assessment or assurance calculations. This has motivated interest in analytical approaches. Fang and Jin\citep{FangJin2021} proposed approximate sample size calculations for estimands with continuous or binary outcomes, and Fang et al.\citep{FangJinWu2024} extended this work to time-to-event endpoints with a single IE. However, these approaches rely on simplifying assumptions for certain strategies. For example, the approach by Fang et al.\citep{FangJinWu2024} for treatment policy and hypothetical handling in estimands with time-to-event endpoints does not explicitly account for the timing of the IE, effectively assuming all IEs occur at randomisation. Such assumptions are likely to result in overly conservative power estimates and inflated sample sizes. 

In this paper, we develop analytical power calculations for estimands with time-to-event endpoints (Section~\ref{sec:methods}). We assume a fixed follow-up period but also demonstrate through simulation that the methods may be applied in settings where follow-up times vary between participants. We initially focus on a single IE handled using a treatment policy, hypothetical, or composite strategy. In contrast to existing approaches, the timing of the IE is explicitly incorporated into the calculations, leading to more accurate estimation of power and required sample size. We then extend the approach to settings with multiple IEs handled using combinations of the three strategies considered. Constant hazards are assumed for all event times in the calculations. In Section~\ref{sec:sim}, simulation studies are used to assess the accuracy of the proposed methods, including settings where design or distributional assumptions are violated. In Section~\ref{sec:case}, we present a case study to illustrate the practical application of the proposed methods and the impact of IEs on trial design, and Section~\ref{sec:discussion} concludes with a discussion.

\section{Methods} \label{sec:methods}

\subsection{General power calculation} \label{sec:general}
 
Consider a trial with two treatment groups, control ($j=0$) and experimental ($j=1$), with $n_j$ participants assigned to treatment group $j$, and in which participants are to be followed for a fixed time $\tau$ after randomisation.

We start in a setting where no IEs are assumed to occur, or in other words, all participants remain on randomised treatment up to the occurrence of the endpoint or to time $\tau$. Denote event times for the endpoint of interest, $Y$, by $T_{ij}^Y$ for participant $i$ on treatment $j$ ($i=1,\ldots, n_j$). We assume that event times $T_{ij}^Y \sim \text{Exp}(\lambda_j)$ are exponentially distributed with rate parameter $\lambda_j$ on treatment $j$, and denote the density and distribution functions of $T_{ij}^Y$ by $f(t;\lambda_j) = \lambda e^{-\lambda_j t}$ and $F(t;\lambda_j) = 1 - e^{-\lambda_j t}$ respectively. We further assume that event times will be analysed using a Cox proportional hazards model to estimate the hazard ratio (HR), $\Delta$.

The maximum (partial) likelihood estimator of the log HR, $\log(\widehat{\Delta})$, is assumed to be normally distributed with mean $\log(\Delta)$ and variance $\sigma^2 = D_0^{-1} + D_1^{-1}$ where $D_0$ and $D_1$ are the expected number of events in the control and experimental treatment groups respectively, by the end of follow-up, $\tau$.\citep{SillRubinstein2009} The power for a two-sided significance level $\alpha$ is
\begin{equation}
\mathrm{Power} = 1 - \Phi\!\left(\left|\frac{\log \Delta}{\sigma}\right| - z_{1-\alpha/2}\right). \label{eq:power}
\end{equation}

To calculate power for an estimand we therefore first require assumptions for $\Delta$ and $D_j$. In the simple case that no IEs are assumed to occur then $\Delta = \lambda_1/\lambda_0$ and $D_j = n_j F(\tau;\lambda_j)$. In the following sections we derive values for $\Delta$ and $D_j$ for estimands with different numbers of IEs and handling strategies.

\subsection{One IE} \label{sec:oneIE}

We first consider estimands with a single intercurrent event $E$ defined, such as treatment discontinuation or use of rescue medication. Denote event times for $E$ by $T_{ij}^E$ for participant $i$ on assigned treatment $j$ ($i=1,\ldots, n_j$). We assume that IE times are exponentially distributed with parameter $\kappa_j$ ($T_{ij}^E \sim \text{Exp}(\kappa_j)$) and that a participant's risk of experiencing the IE is unrelated to their risk of the primary endpoint (in practice such assumptions may be violated to some degree, and we assess deviations to these assumptions in the simulation study), while the effect of the IE on the endpoint is determined by the chosen handling strategy.

In the following section, we derive the value of the estimand, in this case the HR, $\Delta^\mathcal{S}$, under the chosen handling strategy $\mathcal{S}$, along with the expected sampling variance of $\log(\Delta^\mathcal{S})$ for which we derive the corresponding expected number of events in each arm, $D_j^\mathcal{S}$.

\subsubsection{Composite strategy}

Under a composite strategy ($\mathcal{S} = \mathcal{C}$) the IE is defined as part of the outcome measure, and so interest is in time to $E$ or $Y$, whichever occurs first for each participant. The event time under the composite strategy is therefore $T_{ij}^\mathcal{C} = \min(T_{ij}^Y, T_{ij}^E)$. Since the minimum of two independent exponential random variables is also exponential with rate equal to the sum of the individual rates,\citep{Ross2014} the distribution of event times for the composite endpoint is $T_{ij}^\mathcal{C} \sim \text{Exp}(\lambda_j + \kappa_j)$ and the hazard ratio is
\[
\Delta^\mathcal{C} = \frac{\lambda_1 + \kappa_1}{\lambda_0 + \kappa_0}.
\] 
Furthermore, to calculate the variance of $\log \Delta^\mathcal{C}$, the expected number of composite events on treatment arm $j$ by time $\tau$ is $D_j^\mathcal{C} = n_j F(\tau; \lambda_j + \kappa_j)$.

\subsubsection{Hypothetical strategy}

Under a hypothetical strategy ($\mathcal{S} = \mathcal{H}$), it is typical to remove information after the IE in the analysis.\citep{BartlettDaniel2025} We therefore assume that participant $i$ is censored at time $T_{ij}^E$ if it is earlier than $T_{ij}^Y$ so that any $Y$ events occurring after $E$ are ignored or unobserved. The event time under the hypothetical strategy is $T_{ij}^\mathcal{H} = \min(T_{ij}^Y, T_{ij}^E)$, and the event indicator is $Y_{ij}^\mathcal{H} = \mathbb{I}(T_{ij}^Y < T_{ij}^E)$ so that $Y_{ij}^\mathcal{H} = 1$ corresponds to observing $Y$ prior to $E$, and $Y_{ij}^\mathcal{H} = 0$ corresponds to censoring at time $T_{ij}^E$.

Under the assumption of independent censoring at $T_{ij}^E$, the true cause-specific hazard for $Y$ remains $\lambda_j$, and hence the underlying hazard ratio $\Delta^\mathcal{H}$ equals $\Delta$ (the hazard ratio if no IEs occur). Furthermore, under the independent-censoring assumption, the corresponding Cox model estimator of $\Delta^\mathcal{H}$ is consistent for this quantity.\citep{Jackson2014, Mozumder2026, Rufibach2019}

Since participants are censored at $T_{ij}^E$ if it occurs prior to $T_{ij}^Y$, the expected number of observed $Y$ events is reduced. Appendix A shows that the probability $F_j^\mathcal{H}(t)$ of observing $Y$ by time $t$ and before $E$ is
\[
F_j^\mathcal{H}(t) = P\!\left(T_{ij}^\mathcal{H} < t,\, Y_{ij}^\mathcal{H} = 1\right) = \frac{\lambda_j}{\lambda_j + \kappa_j} F(t;\lambda_j + \kappa_j)
\]
Hence, the expected number of observed $Y$ events on treatment arm $j$ by time $\tau$ is $D_j^\mathcal{H} = n_j F_j^\mathcal{H}(\tau)$.

\subsubsection{Treatment Policy strategy}

Under a treatment policy strategy ($\mathcal{S} = \mathcal{P}$), all data observed following $E$ are included in the analysis of $Y$. We assume that occurrence of $E$ modifies the future (post-IE) hazard of $Y$ by a fixed hazard ratio, for instance, due to stopping or changing treatment. Denote by $\lambda_j^*$ the hazard for $Y$ on treatment arm $j$ after the occurrence of $E$.

Given this assumption that the risk of the endpoint $Y$ depends only on whether $E$ has occurred or not, and that the risk of $Y$ is independent of trial time (due to constant hazards), the joint process of $E$ and $Y$ can be represented as a time-homogeneous Markov multistate model, with an intermediate state entered on occurrence of $E$, and cause-specific intensities $\lambda_j$, $\kappa_j$ and $\lambda_j^*$. The setup is analogous to the multistate models for progression-free and overall survival in oncology, and the quantities required below could equivalently be obtained from these intensities directly, following the computations in Meller et al.\citep{Meller2019} and Beyersmann et al.\citep{Beyersmann2012} We instead derive them in terms of the event times $T_{ij}^Y$ and $T_{ij}^E$, with the post-IE hazard $\lambda_j^*$ applying if $T_{ij}^E < T_{ij}^Y$ and $T_{ij}^E$ being irrelevant otherwise. We use this formulation for the joint process because it applies uniformly across the handling strategies and multiple-IE combinations considered in Sections~\ref{sec:oneIE}--\ref{sec:multIE}.

The probability of $Y$ occurring by time $t$ under a treatment policy strategy is the sum of the joint probabilities of its occurrence when $E$ has occurred prior to $Y$ or not:
\[
F_j^\mathcal{P}(t) = P\!\left(T_{ij}^Y \leq t\right) = P\!\left(T_{ij}^Y \leq t,\, T_{ij}^E > t\right) + P\!\left(T_{ij}^Y \leq t,\, T_{ij}^E \leq t\right)
\]

By assuming $T_{ij}^Y$ and $T_{ij}^E$ are independent before $E$ occurs (i.e.\ $Y$ proceeds at its baseline rate $\lambda_j$ unaffected by $E$), the probability of $Y$ occurring prior to $t$ and $E$ occurring after $t$ in a patient is
\begin{align*}
P\!\left(T_{ij}^Y \leq t,\, T_{ij}^E > t\right) &= P\!\left(T_{ij}^Y \leq t \mid T_{ij}^E > t\right) P\!\left(T_{ij}^E > t\right) \\
&= F(t;\lambda_j)\bigl(1 - F(t;\kappa_j)\bigr) \\
&= \bigl(1 - e^{-\lambda_j t}\bigr)e^{-\kappa_j t}
\end{align*}

Given that occurrence of $E$ is assumed to modify the hazard for $Y$, the probability of $Y$ and $E$ occurring prior to $t$ depends on the value of $T_{ij}^E$:
\[
P\!\left(T_{ij}^Y \leq t,\, T_{ij}^E \leq t\right) = \int_0^t P\!\left(T_{ij}^Y \leq t \mid T_{ij}^E = s\right) f(s;\kappa_j)\,ds
\]

Appendix B shows that the probability $P(T_{ij}^Y \leq t \mid T_{ij}^E = s)$ follows a piecewise exponential distribution where the hazard for $Y$ changes from $\lambda_j$ to $\lambda_j^*$ at time $s < t$, and that
\begin{align*}
P\!\left(T_{ij}^Y \leq t,\, T_{ij}^E \leq t\right)
&= \int_0^t \!\bigl(1 - e^{-(\lambda_j s + \lambda_j^*(t-s))}\bigr)\kappa_j e^{-\kappa_j s}\,ds \\
&= 1 - e^{-\kappa_j t} + \frac{\kappa_j}{\lambda_j - \lambda_j^* + \kappa_j}\!\left(e^{-(\lambda_j + \kappa_j)t} - e^{-\lambda_j^* t}\right)
\end{align*}

The distribution function for $Y$ on treatment arm $j$ under the treatment policy strategy, $F_j^\mathcal{P}(t)$, is therefore
\begin{equation}
F_j^\mathcal{P}(t) = 1 - \frac{\lambda_j - \lambda_j^*}{\lambda_j - \lambda_j^* + \kappa_j}e^{-(\lambda_j+\kappa_j)t} - \frac{\kappa_j}{\lambda_j - \lambda_j^* + \kappa_j}e^{-\lambda_j^* t} \label{eq:FjP}
\end{equation}
and the expected number of events on treatment arm $j$ by time $\tau$ is $D_j^\mathcal{P} = n_j F_j^\mathcal{P}(\tau)$.

The hazard ratio $\Delta^\mathcal{P}$ under the treatment policy strategy can be calculated using the following formula from Schemper et al.\citep{Schemper2009} for the Cox model-equivalent average hazard ratio, i.e.\ the quantity to which the Cox partial-likelihood estimator converges asymptotically:
\begin{equation}
\Delta^\mathcal{P} = \frac{\displaystyle\int_0^\tau \frac{h_1^\mathcal{P}(t)}{h_0^\mathcal{P}(t) + h_1^\mathcal{P}(t)}\bigl(f_0^\mathcal{P}(t) + f_1^\mathcal{P}(t)\bigr)\,dt}{\displaystyle\int_0^\tau \frac{h_0^\mathcal{P}(t)}{h_0^\mathcal{P}(t) + h_1^\mathcal{P}(t)}\bigl(f_0^\mathcal{P}(t) + f_1^\mathcal{P}(t)\bigr)\,dt} \label{eq:DeltaP}
\end{equation}
where $h_j^\mathcal{P}(t)$ and $f_j^\mathcal{P}(t)$ are the hazard and density functions for event times on $Y$ in group $j$ and can be calculated using the distribution function in \eqref{eq:FjP}:
\begin{align*}
f_j^\mathcal{P}(t) 
&= \frac{d F_j^\mathcal{P}(t)}{dt} \\
&= \frac{\lambda_j - \lambda_j^*}{\lambda_j - \lambda_j^* + \kappa_j} f(t;\kappa_j + \lambda_j) + \frac{\kappa_j}{\lambda_j - \lambda_j^* + \kappa_j} f(t;\lambda_j^*)
\end{align*}
and
\begin{align*}
h_j^\mathcal{P}(t) 
&= \frac{f_j^\mathcal{P}(t)}{1 - F_j^\mathcal{P}(t)} \\
&= \frac{(\lambda_j + \kappa_j)(\lambda_j - \lambda_j^*) + \lambda_j^*\kappa_j\, e^{-(\lambda_j^* - \lambda_j - \kappa_j)t}}
{(\lambda_j - \lambda_j^*) + \kappa_j\,e^{-(\lambda_j^* - \lambda_j - \kappa_j)t}}
\end{align*}

The HR $\Delta^\mathcal{P}$ can then be computed using \eqref{eq:DeltaP}, for example using numerical integration.

Note that once $\lambda_j \neq \lambda_j^*$, the marginal hazard for $Y$ on each arm becomes a time-varying mixture of the pre- and post-IE hazards, weighted by the (time-dependent) probability of having already experienced $E$. Consequently, the treatment policy strategy induces non-proportional hazards between arms by construction, even though the underlying pre- and post-IE hazards are each constant. The average hazard ratio $\Delta^\mathcal{P}$ as defined in \eqref{eq:DeltaP} is a weighted average of the time-varying true hazard ratio between arms rather than an estimate of a single, constant underlying HR. Its use here is to approximate the expected value of the Cox partial-likelihood estimator when a standard Cox model assuming proportional hazards is fitted to data that do not necessarily satisfy this assumption. Likewise, the formula for $\sigma^2$ in Section~\ref{sec:general} approximates the variance of that estimator rather than the variance under a correctly specified model, but with the expected event counts derived from the true time-varying event distribution. We therefore treat $\Delta^\mathcal{P}$ and the resulting power calculation as a practical approximation to the behaviour of a Cox model routinely applied in this setting, rather than a solution to the non-proportional hazards problem.

For IEs such as discontinuation of assigned treatment, a conservative estimate of power under the treatment policy strategy might be obtained by assuming the hazard for $Y$ on the active arm immediately switches to that on the control arm following occurrence of the IE, while the hazard on the control arm remains unchanged. If $\lambda_0$ and $\lambda_1$ denote the pre-IE hazards on control and active respectively then, under this assumption, $\lambda_0^* = \lambda_0$ on control so that $f_0^\mathcal{P}(t) = f(t;\lambda_0)$ and $h_0^\mathcal{P}(t) = \lambda_0$. On the active arm, $\lambda_1^* = \lambda_0$, and so $f_1^\mathcal{P}(t)$ and $h_1^\mathcal{P}(t)$ can be calculated using the above formula. Alternative values of $\lambda_0^*$ and $\lambda_1^*$ can also be explored to assess the sensitivity of power to assumptions about post-IE outcomes.

\subsection{Multiple IEs} \label{sec:multIE}

We now consider estimands with two IEs, $E_1$ and $E_2$, which are addressed using different handling strategies. As before we assume that a participant's risk of experiencing an IE is unrelated to their risk of other IEs or the primary endpoint, while allowing occurrence of an IE to modify that risk. Denote the event time for IE $E_l$ ($l=1,2$) in subject $i$ on treatment $j$ by $T_{ij}^{E_l}$ and its hazard on treatment arm $j$ by $\kappa_{lj}$.

\subsubsection{Composite and Hypothetical strategies}

Consider an estimand that applies a composite strategy for $E_1$ and a hypothetical strategy for $E_2$ ($\mathcal{S} = \mathcal{C}, \mathcal{H}$). Under this strategy, the event time is $T_{ij}^{\mathcal{C},\mathcal{H}} = \min(T_{ij}^Y, T_{ij}^{E_1}, T_{ij}^{E_2})$ and the event indicator is $Y_{ij}^{\mathcal{C},\mathcal{H}} = \mathbb{I}(\min(T_{ij}^Y, T_{ij}^{E_1}) \leq T_{ij}^{E_2})$ so that $Y_{ij}^{\mathcal{C},\mathcal{H}} = 1$ corresponds to observing the composite endpoint $Y$ or $E_1$, and $Y_{ij}^{\mathcal{C},\mathcal{H}} = 0$ corresponds to censoring at $T_{ij}^{E_2}$.

Assuming $E_2$ induces independent censoring for the composite endpoint $\min(T_{ij}^Y, T_{ij}^{E_1})$, the HR targets the cause-specific HR for $Y$ or $E_1$:
\[
\Delta^{\mathcal{C},\mathcal{H}} = \frac{\lambda_1 + \kappa_{11}}{\lambda_0 + \kappa_{10}}
\]

The expected number of observed (composite) events by time $\tau$ on treatment arm $j$ can then be derived in a similar manner to that shown in Appendix A, so that
\[
D_j^{\mathcal{C},\mathcal{H}} = n_j P\!\left(T_{ij}^{\mathcal{C},\mathcal{H}} < \tau,\, Y_{ij}^{\mathcal{C},\mathcal{H}} = 1\right)
= n_j \frac{\lambda_j + \kappa_{1j}}{\lambda_j + \kappa_{1j} + \kappa_{2j}} F(\tau;\lambda_j + \kappa_{1j} + \kappa_{2j})
\]
where $(\lambda_j + \kappa_{1j})/(\lambda_j + \kappa_{1j} + \kappa_{2j})$ is the probability that $Y$ or $E_1$ occurs before $E_2$.

\subsubsection{Composite and Treatment Policy strategies}

We now consider an estimand that applies a composite strategy for $E_1$ and a treatment policy strategy for $E_2$ ($\mathcal{S} = \mathcal{C}, \mathcal{P}$). Furthermore, we assume that occurrence of $E_2$ could modify the subsequent hazards for $Y$ and $E_1$ by a fixed hazard ratio which may differ between the event types. Let $\lambda_j^*$ and $\kappa_{1j}^*$ denote the hazards for $Y$ and $E_1$ respectively on treatment arm $j$ following occurrence of $E_2$ and let $\nu_j = \lambda_j + \kappa_{1j}$ and $\nu_j^* = \lambda_j^* + \kappa_{1j}^*$ be the hazard for the composite endpoint of $Y$ and $E_1$, pre and post occurrence of $E_2$. Using a similar approach to the treatment policy strategy for a single IE but replacing the hazards for $Y$ ($\lambda_j$ and $\lambda_j^*$) by the corresponding hazards for the composite outcome ($\nu_j$ and $\nu_j^*$), the distribution function is
\[
F_j^{\mathcal{C},\mathcal{P}}(t) = 1 - \frac{\nu_j - \nu_j^*}{\nu_j - \nu_j^* + \kappa_{2j}} e^{-(\nu_j + \kappa_{2j})t} - \frac{\kappa_{2j}}{\nu_j - \nu_j^* + \kappa_{2j}} e^{-\nu_j^* t}
\]

The HR $\Delta^{\mathcal{C},\mathcal{P}}$ can then be calculated in a similar manner to $\Delta^\mathcal{P}$ using formula \eqref{eq:DeltaP}, and the expected number of composite events on arm $j$ is $D_j^{\mathcal{C},\mathcal{P}} = n_j F_j^{\mathcal{C},\mathcal{P}}(\tau)$.

\subsubsection{Hypothetical and Treatment Policy strategies} \label{sec:HP}

Finally, we consider an estimand that applies a hypothetical strategy for $E_1$ and a treatment policy strategy for $E_2$ ($\mathcal{S} = \mathcal{H}, \mathcal{P}$). As in Section 2.3.2 we assume that occurrence of $E_2$ could alter the hazard for $Y$ and/or $E_1$ and denote the corresponding post-$E_2$ hazards by $\lambda_j^*$ and $\kappa_{1j}^*$ respectively. The event time for each participant on arm $j$ is $T_{ij}^{\mathcal{H},\mathcal{P}} = \min(T_{ij}^Y, T_{ij}^{E_1})$ and the endpoint indicator is $Y_{ij}^{\mathcal{H},\mathcal{P}} = \mathbb{I}(T_{ij}^Y \leq T_{ij}^{E_1})$.

Assuming that occurrence of $E_1$ creates an independent censoring process for $Y$, the HR is the same as that for an estimand with a single IE using a treatment policy strategy, i.e.\ $\Delta^{\mathcal{H},\mathcal{P}} = \Delta^\mathcal{P}$. However, due to the censoring by $E_1$, the expected number of observed $Y$ events is reduced. In a similar manner to the treatment policy case (Section 2.2.3), the probability of observing $Y$ before both time $t$ and $E_1$ on arm $j$, $F_j^{\mathcal{H},\mathcal{P}}(t)$, can be determined by noting:
\begin{align*}
F_j^{\mathcal{H},\mathcal{P}}(t) &= P\!\left(T_{ij}^{\mathcal{H},\mathcal{P}} \leq t,\, Y_{ij}^{\mathcal{H},\mathcal{P}} = 1\right) \\
&= P\!\left(T_{ij}^Y \leq t,\, T_{ij}^Y \leq T_{ij}^{E_1},\, T_{ij}^{E_2} > t\right) + P\!\left(T_{ij}^Y \leq t,\, T_{ij}^Y \leq T_{ij}^{E_1},\, T_{ij}^{E_2} \leq t\right)
\end{align*}

As in the previous section, letting $\nu_j = \lambda_j + \kappa_{1j}$ and $\nu_j^* = \lambda_j^* + \kappa_{1j}^*$ denote the hazard for $\min(T_{ij}^Y, T_{ij}^{E_1})$ pre and post occurrence of $E_2$ respectively, Appendix C shows that
\[
F_j^{\mathcal{H},\mathcal{P}}(t) = \frac{\lambda_j}{\nu_j + \kappa_{2j}}\!\left(1 - e^{-(\nu_j + \kappa_{2j})t}\right)
+ \frac{\lambda_j^*}{\nu_j^*}\kappa_{2j}\!\left[\frac{1 - e^{-(\nu_j + \kappa_{2j})t}}{\nu_j + \kappa_{2j}} - e^{-\nu_j^* t}\frac{1 - e^{-(\nu_j - \nu_j^* + \kappa_{2j})t}}{\nu_j - \nu_j^* + \kappa_{2j}}\right]
\]

The expected number of events on arm $j$ by time $\tau$ is then $D_j^{\mathcal{H},\mathcal{P}} = n_j F_j^{\mathcal{H},\mathcal{P}}(\tau)$.

Note that if $E_2$ is an IE such as treatment discontinuation and we assume that its occurrence has no impact on the hazards for $Y$ and $E_1$, which may be plausible for an inactive control such as placebo, then $\nu_j = \nu_j^*$. In this instance the above formula reduces to $F_j^{\mathcal{H},\mathcal{P}}(t) = \frac{\lambda_j}{\nu_j}(1 - e^{-\nu_j t})$ which equals the formula for $F_j^\mathcal{H}(t)$ in Section 2.2.2, and is as expected given that the risk of the endpoint is assumed to be unaffected by $E_2$.

\subsubsection{Three IEs}

Suppose an estimand has three IEs $E_1$, $E_2$ and $E_3$ in which a composite strategy is used for $E_1$, a hypothetical strategy is used for $E_2$ and a treatment policy strategy is used for $E_3$ ($\mathcal{S} = \mathcal{C}, \mathcal{H}, \mathcal{P}$). A similar approach to that described in Section~\ref{sec:HP} can be used to determine the function $F_j^{\mathcal{C},\mathcal{H},\mathcal{P}}(t)$ by replacing the hazard for $Y$ by the corresponding hazard for the composite endpoint ($Y$ or $E_1$). Further details are provided in Appendix D.

\section{Simulation Study} \label{sec:sim}

To assess the accuracy of the proposed approaches we compared calculated power values to those obtained through simulations of patient-level data. These comparisons were made across a variety of estimands, each with a different number of IEs or set of handling strategies. Specifically, we considered the following scenarios:
\begin{enumerate}
  \item One IE, with IE times independent of pre-IE event times for outcome $Y$
  \item One IE, with IE times correlated with pre-IE event times for outcome $Y$
  \item Two IEs, with IE times independent of each other and pre-IE event times for outcome $Y$ (see Appendix F)
\end{enumerate}

In the simulations, we varied several key design parameters, including sample size (consistent with phase 2 and 3 trials), control arm event rates, hazard ratios (both directions of effect), and IE rates. Generally, we considered a fixed follow-up duration and assumed constant hazards for events except for scenario 1, where we also separately considered variable follow-up durations and non-constant hazards for both the outcome $Y$ and IE $E$ to assess deviations to calculated power when these design or distributional assumptions are violated.

\subsection{One IE} \label{sec:simOneIE}

The first simulation study assessed estimands with a single IE, $E$, assumed to be independent of outcome $Y$, with the following strategies:
\begin{itemize}
  \item Composite ($\mathcal{C}$): time to the earliest $E$ or $Y$ event
  \item Hypothetical ($\mathcal{H}$): time to event $Y$ ignoring data following occurrence of $E$
  \item Treatment policy ($\mathcal{P}$): time to event $Y$ for which the hazard for $Y$ on active is assumed to switch to that on placebo following occurrence of $E$. The hazard remains unaffected on placebo.
\end{itemize}

\subsubsection{Constant and non-constant hazards}

Data were simulated for a randomised trial of two treatment arms, active ($j=1$) and control ($j=0$), with $n$ participants per arm. Event times for $Y$, $T_{ij}^Y$, and the IE, $T_{ij}^E$, were drawn from independent Weibull distributions with scale parameters $\lambda_j^Y$ and $\lambda_j^E$, and shape parameters $b_j^Y$ and $b_j^E$ respectively, for treatment $j$ ($j = 0, 1$ and $i = 1,\ldots,n$). All combinations of the following parameters were assessed for each IE strategy:
\begin{enumerate}
  \item Sample size per arm: $n \in \{100, 250, 500\}$.
  \item Control arm scale parameter for $Y$: $\lambda_0 = -\log(1-p_0)/\tau^{b^Y}$, where $p_0 \in \{0.25, 0.50, 0.75\}$ is the risk of $Y$ by time $\tau$ on control.
  \item Active arm scale parameter for $Y$: $\lambda_1 = \lambda_0 \Delta^Y$, where hazard ratio $\Delta^Y \in \{0.667, 0.8, 1.25, 1.5\}$. $\Delta^Y > 1$ corresponds to outcomes for which a shorter time to $Y$ is more favourable, e.g.\ time to cure.
  \item Weibull shape parameter for $Y$ (same for both treatment arms): $b^Y \in \{0.67, 1, 1.5\}$, i.e.\ decreasing, constant and increasing hazards.
  \item Control arm scale parameter for $E$: $\kappa_0 = -\log(1-q_0)/\tau^{b^E}$ where $q_0 = \pi_0 p_0$ and $\pi_0 \in \{0.1, 0.25, 0.5\}$, i.e., the risk of the IE by time $\tau$ on arm $j=0$ is a proportion of the corresponding risk for $Y$.
  \item Active arm scale parameter for $E$: $\kappa_1 = \kappa_0 \Delta^E$ where $\Delta^E \in \{0.75, 1, 1.25\}$ is the HR for the IE.
  \item Weibull shape parameter for $E$ (same for both treatment arms): $b^E \in \{0.67, 1, 1.5\}$, i.e.\ decreasing, constant and increasing hazards.
\end{enumerate}

Under the treatment policy strategy, participants on the active arm with $T_{i1}^E < T_{i1}^Y$ were assumed to switch their hazard for $Y$ to that of the placebo arm at time $T_{i1}^E$ and follow the corresponding Weibull hazard curve for the placebo arm thereafter.

In total 2916 design scenarios were assessed for each IE strategy, covering a wide range of powers. In addition, designs in which no IEs were assumed to occur were also included (i.e.\ $\kappa_0 = 0$) to determine whether any differences between calculated and simulated power for the various handling strategies could be due to formula \eqref{eq:power} for calculating power.

We initially used a fixed follow-up time of $\tau = 1$. The event times and event indicators for each strategy are summarised in Table~\ref{tab:table1}.

\begin{table}[H]
\centering
\caption{Definition of event times and indicators for each IE strategy assessed in simulations of estimands with one IE, independent of outcome $Y$, or no IEs.}
\label{tab:table1}
\begin{tabular}{p{3cm} p{5.5cm} p{5.5cm}}
\toprule
\textbf{Strategy,} $\mathcal{S}$ & \textbf{Event Time,} $T_{ij}^{\mathcal{S}}$ & \textbf{Event Indicator by time} $\tau$, $Y_{ij}^{\mathcal{S}}$ \\
\midrule
No IEs & $\min(T_{ij}^Y, \tau)$ & $\mathbb{I}(T_{ij}^Y < \tau)$ \\[6pt]
Composite, $\mathcal{C}$ & $\min(T_{ij}^E, T_{ij}^Y, \tau)$ & $\mathbb{I}(T_{ij}^{\mathcal{C}} < \tau)$ \\[6pt]
Hypothetical, $\mathcal{H}$ & $\min(T_{ij}^E, T_{ij}^Y, \tau)$ & $\mathbb{I}(T_{ij}^Y < \min(T_{ij}^E, \tau))$ \\[6pt]
\multirow{4}{*}{Treatment Policy, $\mathcal{P}$}
& \(\begin{aligned}
j=0:\quad & \min(T_{i0}^Y, \tau)
\end{aligned}\)
& \multirow{4}{*}{$\mathbb{I}(T_{ij}^{\mathcal{P}} < \tau)$} \\

& \(\begin{aligned}
j=1:\quad & \min(T_{i1}^Y, \tau), \text{ if } T_{i1}^Y < T_{i1}^E \\
          & \min(T_{i1}^E + T_{i1}^{Y^*}, \tau), \text{ otherwise}
\end{aligned}\)
& \\
\bottomrule
\end{tabular}
\end{table}

Event times were analysed using a Cox proportional hazards model. Power was estimated as the proportion of simulated trials in which the two-sided p-value was $<0.05$ for the HR of the effect of treatment $j=1$ vs $j=0$. For each scenario, 10,000 simulations were used to ensure the Monte Carlo standard error (MCSE) of the power estimate (maximised at 50\% power) is less than 0.5\%.

\begin{figure}[H]
\centering
\includegraphics[width=0.95\textwidth]{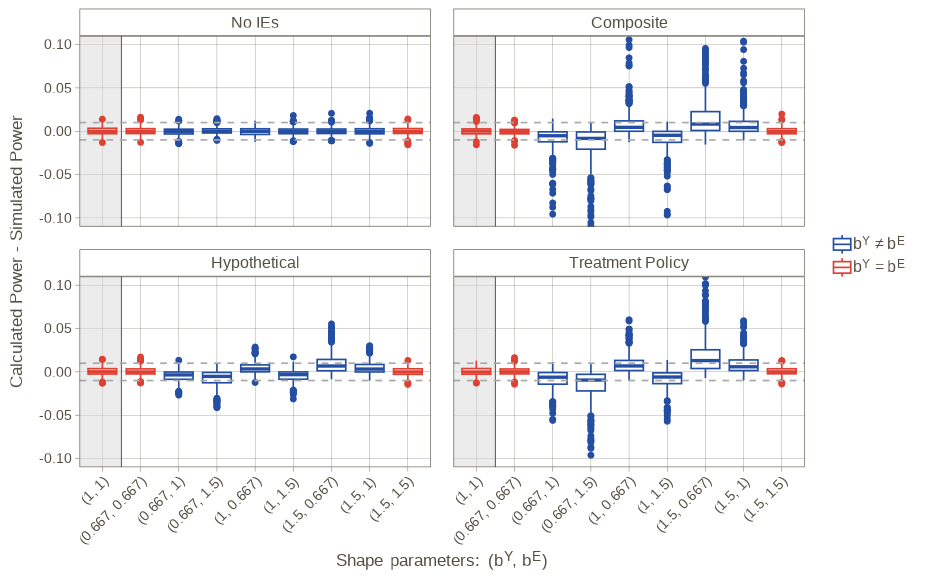}
\caption{Box plots of differences between calculated and simulated power for each pair of Weibull shape parameters $(b^Y, b^E)$ and IE strategy considered. Dashed lines indicate the expected 95\% CIs for the power difference under MC error.}
\label{fig:fig1}
\end{figure}

Figure~\ref{fig:fig1} shows a summary of the differences between calculated and simulated power for each combination of shape parameters $(b^Y, b^E)$ and each IE strategy considered. For the base-case scenario which aligns with assumptions made in the calculations, i.e.\ $(b^Y, b^E) = (1, 1)$ -- exponentially distributed event times, estimates were within MC error (i.e.\ predominantly up to $\pm 1\%$ difference) for all IE strategies. Some slightly larger differences were apparent for designs with lower levels of power and the smallest sample size considered ($n=100$), see Figure~\ref{fig:A1} in Appendix E. These were scenarios in which a small number of events were observed, where standard Cox partial likelihood estimation can exhibit appreciable finite sample bias and may suffer from monotone likelihood leading to unstable estimates.

With no IEs (i.e.\ all patients are assumed to remain on randomised treatment), calculated and simulated power agreed up to MC error for all values of $b^Y$ ($b^E$ is not applicable here). This is expected, since power depends only on the number of events by time $\tau$, and the scale parameters were chosen to give a consistent event rate by $\tau$ across all shape parameters. In the presence of an IE, agreement depended on the shapes of the hazard curves and the estimand strategy. When $b^Y = b^E$, differences were within MC error for every strategy. Discrepancies emerged only when $b^Y \neq b^E$, and increased with the magnitude of the difference between the two shape parameters. These discrepancies were modest under the hypothetical strategy (at most approximately $\pm 0.05$), and larger under the treatment policy and composite strategies.

\subsubsection{Variable follow-up durations}

We next considered trials with variable follow-up durations, which may be more commonly used in practice. A maximum trial duration of $t_{\max} = 1$ was imposed and patients were assumed to be recruited at a constant rate up to time $t_{\text{rec}} \in \{0, 0.25, 0.5\}$; $t_{\text{rec}} = 0$ recovers the fixed follow-up case ($\tau = 1$ for all patients) of the previous section and serves as a comparator. Enrolment times were simulated as $t_{ij} \sim U[0, t_{\text{rec}}]$, giving a follow-up time of $\tau_{ij} = t_{\max} - t_{ij}$ for patient $i$ on arm $j$. Event times were simulated as described in the previous section using the parameters in Table~\ref{tab:table1}, with $\tau$ replaced by $\tau_{ij}$, and were restricted to the case of independent exponentially distributed times for $Y$ and $E$ (324 design scenarios per accrual rate). Power was calculated using the expected follow-up time across participants: since accrual is linear, this is $t_{\max} - t_{\text{rec}}/2$, e.g.\ 0.875 when $t_{\text{rec}} = 0.25$.

\begin{figure}[H]
\centering
\includegraphics[width=0.95\textwidth]{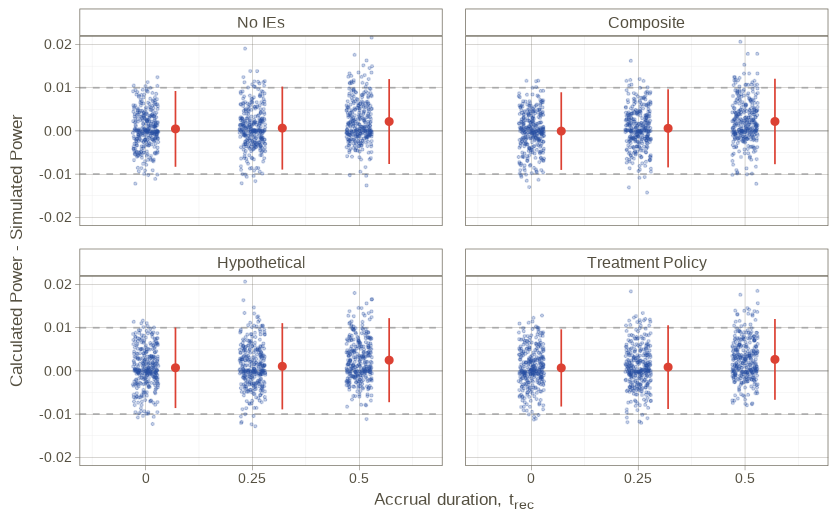}
\caption{Difference between calculated and simulated power for different accrual durations and IE strategies. Circle marker is the mean difference with 95\% reference range for each scenario considered. Dashed lines show the expected 95\% ranges under MC error. Accrual duration of 0 corresponds to fixed follow-up of all patients to time $\tau = 1$.}
\label{fig:fig2}
\end{figure}

Figure~\ref{fig:fig2} shows the difference between calculated and simulated power by accrual duration and estimand strategy. Calculated and simulated power generally agreed to within MC error throughout, although a small positive bias emerged as accrual duration, and hence the variability in follow-up times, increased.

\subsection{One IE (Dependent Event Times)}

We also investigated scenarios where event times for the endpoint $Y$ and a single IE, $E$, are exponentially distributed and dependent to assess how power deviates from the calculated values (which assume independence between $Y$ and $E$) across the IE strategies considered. Such dependencies may be clinically relevant; for instance, participants experiencing poorer treatment outcomes might be more prone to IEs such as treatment discontinuation or rescue medication use.

We considered scenarios where an unfavourable event (e.g.\ death or hospitalisation) is assessed and so a HR of $\Delta^Y < 1$ implies benefit. In this case occurrence times for $E$ and $Y$ may, to some degree, be positively correlated (i.e.\ patients with a shorter event time for the primary endpoint are also likely to have a shorter event time for the IE). To simulate correlated event times, we used a Gaussian copula, which assumes a positive, symmetric dependence structure across events times. Dependence between event times was quantified using the median concordance probability ($\widetilde{\mathrm{CP}}$), defined as the probability that event times are either both above or both below their respective median times.

The dependence parameter, $\theta$, for the Gaussian copula was derived from $\widetilde{\mathrm{CP}}$ via its relation to Blomqvist's beta ($\beta = 2\widetilde{\mathrm{CP}} - 1$) using the formula $\theta = \sin(\pi\beta/2)$.\citep{Joe2015} To simulate correlated event times for each participant, uniformly distributed values, $U_{ij}^Y$ and $U_{ij}^E$, were first generated via the copula. For the treatment policy strategy, a 3-dimensional copula was used to generate a third uniform variable, $U_{ij}^*$, to derive post-IE event times for participants on treatment arm $j=1$. Event times for $Y$, $E$, and the post-IE event time $Y^*$ were then calculated as
\begin{itemize}
  \item $T_{ij}^Y = -\log(U_{ij}^Y)/\lambda_j$
  \item $T_{ij}^E = -\log(U_{ij}^E)/\kappa_j$
  \item $T_{i1}^* = -\log(U_{i1}^*)/\lambda_0$
\end{itemize}

Throughout the simulations, dependence was assumed to be exchangeable across all event types. The principles detailed in Table~\ref{tab:table1} for deriving event times and corresponding indicators for each handling strategy were used.

To investigate the impact of varying dependence levels between event times on power relative to the calculated value (which assumes independence between event times), we focused on a specific design scenario from Section~\ref{sec:simOneIE}. The scenario of interest used $n=500$, $\Delta^Y=0.8$ and $p_0=0.75$ which, under the assumption that no IEs occur, has 84\% power. For the IE, we assume $\pi_0=0.25$, corresponding to an IE rate of 0.1875 by time $\tau$ on the control arm ($j=0$). On the active arm, we assessed IE HRs of $\Delta^E=1$ (no effect on the IE rate), $\Delta^E=0.75$ (active treatment reduces the IE rate vs control) and $\Delta^E=1.25$ (active treatment increases the IE rate vs control). We explored both moderate ($\widetilde{\mathrm{CP}}=0.6$) and high ($\widetilde{\mathrm{CP}}=0.8$) levels of median concordance (dependence) between $Y$ and $E$.

Table~\ref{tab:table2} summarises the differences between calculated and simulated power for the various handling strategies and levels of dependence considered, within the design scenario of interest. A key observation in the simulations is that when events times for $Y$ and $E$ are positively correlated, fewer IEs are observed in the simulations than would be anticipated under independence. This occurs because positive correlation increases the probability of both events ($Y$ and $E$) occurring in the same individuals. Furthermore, given the hazard for $Y$ is assumed to be greater than that for $E$, the $Y$ event is more likely to precede $E$ in those individuals who would experience both event types before $\tau$, effectively reducing the observed IE count. The implications of this for each handling strategy, in the design scenario considered in Table~\ref{tab:table2}, were as follows:
\begin{itemize}
  \item \textbf{Composite strategy}: power is higher in the simulations when there is a weaker beneficial effect on the IE than on $Y$ ($\Delta^E > \Delta^Y$). This is because a larger proportion of the composite events in the simulations consists of $Y$ events (on which there is a treatment effect, $\Delta^Y$) rather than $E$ events. The difference in power becomes larger as the difference between $\Delta^E$ and $\Delta^Y$, or the dependence parameter $\widetilde{\mathrm{CP}}$, increases. Conversely, when the treatment's effect on $E$ is stronger than on $Y$ ($\Delta^E=0.75$), the opposite effect is observed.
  \item \textbf{Hypothetical strategy}: power tends to be higher in the simulations since $Y$ events are less likely to be censored by the IE due to the positive correlation. The differences in power are smaller for lower values of $\Delta^E$ because the discrepancy between predicted and simulated events counts are less pronounced in this case due to the lower hazard for the IE on the active arm (i.e.\ fewer $Y$ events are censored).
  \item \textbf{Treatment policy strategy}: simulated power is consistently higher than the calculated values, and this difference increases with higher dependence. Since fewer IEs occur prior to $Y$ when the dependence between these events increases, less post-IE time is also observed, during which period the hazards on $Y$ were assumed equal in both treatment arms. Consequently, the overall HR in the simulations becomes less affected by post-IE outcomes. The differences in power are smaller for lower values of $\Delta^E$ because the IE is less likely to occur on the active arm ($j=1$), thereby having less impact. Differences in power should also be lower when assuming that the impact of the IE on post-IE outcomes is not as severe as that assumed here (i.e.\ instant loss of effect), although this requires further investigation.
\end{itemize}

\begin{table}[H]
\centering
\caption{Difference in calculated power ($^*$assuming independence between the IE and outcome) and simulated power assuming event times for the IE and outcome are weakly ($\widetilde{\mathrm{CP}}=0.6$) or strongly ($\widetilde{\mathrm{CP}}=0.8$) dependent, by IE strategy. Design scenario: $n=500$, $\Delta=0.8$, $p_0=0.75$ and $\pi_0=0.25$. $\widetilde{\mathrm{CP}}$ = median concordance probability between event times on $Y$ and $E$; $\Delta^E$ = HR on IE.}
\label{tab:table2}
\begin{tabular}{lccccc}
\toprule
\textbf{Strategy} & $\boldsymbol{\Delta^E}$ & \shortstack{\textbf{Power}\\\textbf{(Calculated$^*$)}} & $\boldsymbol{\widetilde{\mathrm{CP}}}$ &\shortstack{\textbf{Power}\\\textbf{(Simulated)}} & \shortstack{\textbf{Difference in Power}\\\textbf{(Calculated -- Simulated)}} \\
\midrule
\multirow{6}{*}{Composite}     & \multirow{2}{*}{0.75} & \multirow{2}{*}{0.888} & 0.6 & 0.858 & $0.031$  \\
                                &                       &                        & 0.8 & 0.845 & $0.043$  \\
                                & \multirow{2}{*}{1}    & \multirow{2}{*}{0.752} & 0.6 & 0.762 & $-0.010$ \\
                                &                       &                        & 0.8 & 0.825 & $-0.074$ \\
                                & \multirow{2}{*}{1.25} & \multirow{2}{*}{0.562} & 0.6 & 0.637 & $-0.076$ \\
                                &                       &                        & 0.8 & 0.813 & $-0.251$ \\
\midrule
\multirow{6}{*}{Hypothetical}  & \multirow{2}{*}{0.75} & \multirow{2}{*}{0.817} & 0.6 & 0.810 & $0.007$  \\
                                &                       &                        & 0.8 & 0.839 & $-0.022$ \\
                                & \multirow{2}{*}{1}    & \multirow{2}{*}{0.813} & 0.6 & 0.830 & $-0.017$ \\
                                &                       &                        & 0.8 & 0.862 & $-0.049$ \\
                                & \multirow{2}{*}{1.25} & \multirow{2}{*}{0.809} & 0.6 & 0.850 & $-0.041$ \\
                                &                       &                        & 0.8 & 0.900 & $-0.091$ \\
\midrule
\multirow{6}{*}{Treatment Policy} & \multirow{2}{*}{0.75} & \multirow{2}{*}{0.798} & 0.6 & 0.824 & $-0.026$ \\
                                &                       &                        & 0.8 & 0.850 & $-0.052$ \\
                                & \multirow{2}{*}{1}    & \multirow{2}{*}{0.782} & 0.6 & 0.814 & $-0.032$ \\
                                &                       &                        & 0.8 & 0.845 & $-0.063$ \\
                                & \multirow{2}{*}{1.25} & \multirow{2}{*}{0.766} & 0.6 & 0.810 & $-0.045$ \\
                                &                       &                        & 0.8 & 0.857 & $-0.091$ \\
\bottomrule
\end{tabular}
\end{table}

These simulations demonstrate that the effect of the dependency between event times for the endpoint, $Y$, and an IE, $E$, on power can be complex and likely to differ depending on the specific design scenario, the assumed level of dependency between events, the IE rates, and the selected handling strategy. Some consistent patterns can be deduced based on the scenario explored above, in which the positive dependence between event times and higher hazard for $Y$ means that fewer IEs will be observed compared to when events are independent.

We have not considered scenarios in which the outcome of interest is favourable (e.g.\ time to cure or time to discharge), and so $\Delta^Y > 1$ would be targeted. In this case it is more likely that event times would be negatively correlated with IE times since IEs tend to be unfavourable events (i.e.\ patients with a shorter event time for the primary endpoint are likely to have a longer event time for the IE). This type of dependence may have a different impact on power compared to that in the case of positive dependence. We have also not considered estimands with two dependent IEs, which may also be correlated with events times on $Y$. If there is a strong belief that event times are likely to be dependent, and that this dependency can be quantified, then we recommend running simulations to more accurately assess power changes to assess sensitivity to the assumed level of dependence.

\subsection{Two IEs}

Appendix F presents a simulation study comparing calculated and simulated power for estimands with two IEs, $E_1$ and $E_2$, with event times drawn from independent exponential distributions. The following combination of strategies are considered:
\begin{enumerate}
  \item[(a)] Composite ($E_1$) and hypothetical ($E_2$)
  \item[(b)] Composite ($E_1$) and treatment policy ($E_2$)
  \item[(c)] Hypothetical ($E_1$) and treatment policy ($E_2$)
\end{enumerate}

As with the simulations of estimands with a single IE with exponentially distributed event times, estimates agreed up to MC error and there were no systematic differences identified across any of the design parameters considered.

\section{Case Study} \label{sec:case}

We illustrate the proposed methods using the SYNAPSE trial, which compared mepolizumab to placebo in treating chronic rhinosinusitis with nasal polyps (ClinicalTrials.gov identifier: NCT03085797).\citep{Han2021} Nasal polyposis is a chronic inflammatory condition of the nasal mucosa marked by soft tissue growth in the upper nasal cavity. These polyps can lead to long-term symptoms---including significant nasal obstruction, post-nasal drip, loss of smell, facial pain/pressure, and nasal discharge---that can greatly diminish a participant's quality of life. In severe cases, these issues may ultimately necessitate surgery to remove the polyps, thereby restoring proper aeration of the nasal passages and sinuses.

In the SYNAPSE trial, 200 participants were randomised to each of the mepolizumab and placebo arms and were followed up for a fixed 52-week period. The trial was powered for co-primary endpoints of nasal polyps score and nasal obstruction visual analogue scale (VAS) score and a key secondary endpoint of time to nasal surgery. Since the trial was designed before the implementation of ICH E9 (R1), an estimand for each key endpoint was not formally defined. Therefore, we reconstruct the estimand for time to surgery based on the design of the trial and the original analysis of the endpoint.

Treatment discontinuation was the only IE we identified for the endpoint of time to surgery. The study protocol mandated the continued collection of data from participants who prematurely stopped randomised treatment, and these data were included in the analysis. This is consistent with a treatment policy strategy for the IE of treatment discontinuation. No additional IEs for this endpoint were identified. The summary measure was a HR comparing mepolizumab to placebo.

For the original sample size calculation of the time to nasal surgery endpoint, the assumed true proportions of participants undergoing surgery by week 52 were 40\% for the placebo group and 25\% for the mepolizumab group. Although the study protocol does not clarify whether these rates considered treatment discontinuation and its subsequent effects on surgery rates, we assume they represent the true surgical rates if participants adhered to the treatment for the full trial duration. Assuming that surgery times are exponentially distributed, these rates correspond to hazards of $\lambda_0 = 0.511$ on placebo and $\lambda_1 = 0.288$ on mepolizumab, yielding a hazard ratio of $\Delta = 0.563$. Under these assumptions, a sample size of $n=200$ per arm is estimated to provide roughly 90\% power at the two-sided 5\% significance level.

During the trial, 34 out of 201 participants (17\%) in the placebo group and 23 out of 206 participants (11\%) in the mepolizumab group discontinued study treatment before the final week 52 visit (see Figure 1 of Han et al.\citep{Han2021}). Assuming an exponential distribution for time to treatment discontinuation, these rates correspond to hazards of $\kappa_0 = 0.185$ on placebo and $\kappa_1 = 0.118$ on active treatment. Taking these to be the true IE hazards and using the approach described in Section 2.2.3, the power is calculated to be 85\% for the time to surgery endpoint under the assumption that the hazard for surgery immediately reverts to $\lambda_0$ following treatment discontinuation on the active arm (with no change in hazard on placebo). To recover the original level of power, a sample size increase to $n=225$ per arm (an additional 50 participants total) would be required.

Since mepolizumab treatment effectively reduces nasal polyp size and thus the need for surgery, it is unlikely that the hazard for surgery would immediately match that of the placebo group following treatment discontinuation. In other words, the assumption of an instant loss of treatment effect may be overly conservative in this setting. As a sensitivity analysis, we assumed that the post-IE hazard for surgery on the active arm would be the average of the on-treatment hazards across arms, i.e., $\lambda_1^* = (\lambda_1 + \lambda_0)/2$, to account for a possible carryover effect of the active treatment. Under this assumption, the power is approximately 87\%, and a sample size of $n=213$ per arm would be needed to maintain the original power level.

\begin{figure}[H]
\centering 
\includegraphics[width=0.95\textwidth]{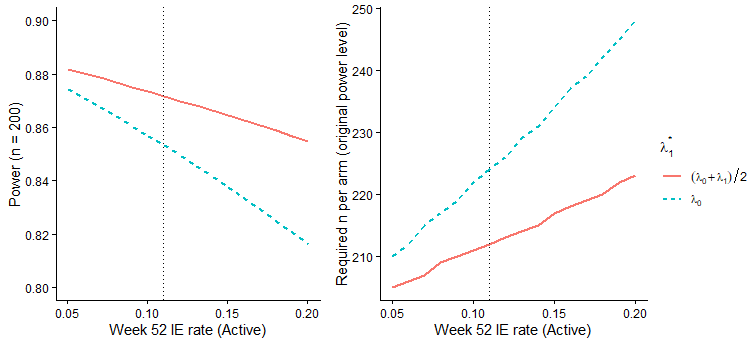}
\caption{Power given $n=200$ and required $n$ to maintain original power for the time to surgery endpoint in SYNAPSE, accounting for impact of different rates of treatment discontinuation and impact on subsequent outcomes on the active arm. $\lambda_1^*$ denotes the assumed hazard rate for surgery following treatment discontinuation. The vertical dashed line indicates the original assumed rate of treatment discontinuation.}
\label{fig:fig3}
\end{figure}

Finally, we evaluated the sensitivity of power for the time to surgery estimand under various treatment discontinuation rates on the active arm, considering rates between 5\% to 20\% by the end of follow-up, and across the two assumed post-IE hazard rates for surgery. As shown in Figure~\ref{fig:fig3}, if the discontinuation rate on the active arm is as high as 20\%, the power could drop to 82\% under the immediate loss of effect assumption, potentially requiring an increase of up to 50 participants per arm compared to the assumption of no treatment discontinuations.

This case study shows that the assumptions about IE rates and, for the treatment policy strategy, post-IE outcomes can have a large impact on sample size requirements and should therefore be carefully considered alongside exploring sensitivity to assumptions over a range of clinically plausible scenarios.

The proposed approach also facilitates assurance calculation to incorporate uncertainty in design parameter values. In this example, prior distributions, potentially informed by formal prior elicitation, may be specified for treatment discontinuation rates and their impact on subsequent outcomes, as well as for additional design parameters such as the hazard ratio. Parameter values can then be sampled from the specified priors and power evaluated for each draw. The resulting values can then be averaged to obtain an overall estimate of the probability of success.

\section{Discussion} \label{sec:discussion}

The ICH E9(R1) addendum states that estimands should inform trial design, particularly sample size calculations.\citep{ICH_E9R1_2019} To achieve this, the impact of IEs and their handling strategies on power should be investigated to ensure the chosen sample size is appropriate. In this paper, we have derived formulae for calculating power for estimands with a time-to-event endpoint, specifically addressing IEs handled using composite, hypothetical, or treatment policy strategies. This work advances previous methods, such as those by Fang et al.\citep{FangJinWu2024}, by explicitly incorporating the timing of IEs to estimate power more accurately, and providing approaches for estimands with multiple IEs.

Our proposed approach does not consider principal stratum or while on treatment strategies since these appear to be much less prevalent in practice.\citep{Bergman2025, Meszaros2024} Fang et al.\citep{FangJinWu2024} propose a basic approximation for these strategies. For while-on-treatment, they assert that occurrence of the IE acts as a form of censoring in which participants are assumed to subsequently stop their assigned treatment after the IE. However, this is not in-line with the while on-treatment strategy which, by definition, focuses solely on the time up to the IE only, thus rendering the time following the IE as irrelevant. Furthermore, their approach to the principal stratum strategy, which involves rescaling sample size based on the proportion of participants not experiencing an IE, assumes that the participants who would experience the IE on each treatment arm are non-overlapping (i.e.\ no participant would experience the IE on both treatment arms). This is unlikely to be the case in practice and may therefore be overly conservative, leading to inflated sample sizes. More rigorous calculations for these two strategies are warranted should their practical application become more widespread.

We conducted a comprehensive simulation study to assess the accuracy of the proposed formulae across various scenarios and assessed deviations to calculated power values when distributional or design assumptions are violated. The formulae demonstrated strong agreement with simulated power under the assumption of independence between event times on the IEs and the primary outcome and when event times are assumed to be both exponentially distributed or follow Weibull distributions with similar shape parameters, otherwise differences in power become apparent particularly for the composite and treatment policy strategies. Extending the methods to assumptions of non-constant hazards or dependent event times is therefore an important area of future research. The proposed methods may provide an initial impression of power and allow a large design space to be explored relatively quickly, but simulations may be required where there is strong prior evidence of substantially different hazard shapes for the outcome and the IE, or of appreciable dependence between them, to obtain a more accurate power estimate under the anticipated data-generating mechanism.

The assumption of independence between event times on the IE and outcome (pre-IE) may not always hold in practice; for example, participants at higher risk of the primary endpoint might be more prone to discontinue treatment early, thereby linking this potential IE with the outcome. Our simulations showed that deviations between calculated and simulated power are directly influenced by the level of dependence between events, with the magnitude of these deviations affected by IE rates, and the chosen handling strategy. For lower levels of dependence or infrequent IEs, independence-based power calculations are likely to remain reasonably accurate, typically within a few percentage points. However, these differences may become more substantial with stronger dependence or more common IEs. Future work is required to extend the proposed methods to scenarios where outcomes and IEs are dependent. For composite handling, a practical simplification would be to model the composite event as a single endpoint and assume a hazard for this (potentially facilitated by existing trial data), which could offer a more accurate power estimate than assuming separate hazards for the IE and outcome under independence. For other handling strategies, similar straightforward workarounds for dependence are less apparent.

The actual extent of the association between IE and outcome event times in clinical settings is unclear. Leveraging existing trial data for a specific disease area could provide valuable empirical insights to inform assumptions regarding IE rates, dependence patterns, and the impact of IEs on post-IE outcomes within a treatment policy framework. For instance, assuming post-IE hazards immediately revert to those on the control arm (instant loss of treatment effect), might be overly conservative in some contexts, as explained in our nasal polyps case study where the impact of treatment discontinuation on subsequent surgery rates is unclear. Employing robust data-driven assumptions, whether for direct power calculations or guiding simulations, is crucial for more accurately aligning sample size with the estimand.

If more than three IEs are defined using the strategies covered in this work there will be at least two IEs handled with the same strategy. In this case, the hazards for the IEs with corresponding strategies can be summed to obtain the hazard for either IE occurring first. The approaches described in this paper can then be applied for the relevant combination of strategies which are to be used. Note that this assumes that all IEs handled using the treatment policy strategy modify the hazards for all other IEs and $Y$ in the same way which might not always be the case in practice. For instance, if the IEs are treatment discontinuation and rescue medication, then it is plausible that treatment discontinuation will increase the risk of an unfavourable endpoint while using rescue medication may reduce it. Extending the methods to such a complex setting is beyond the scope of this paper. Moreover, as the number of IEs considered increases, so does the number of assumptions used. This not only makes the power calculation assumption-heavy but also can complicate sensitivity analysis. In practice we recommend focussing only on those IEs which are likely to have the largest influence on power, to reduce the number of assumptions required.

Our work primarily assumes the HR as the summary measure of the estimand. While HRs are widely used and accepted by regulatory bodies,\citep{FDA2026} their interpretation can be complicated by issues such as the potential for treatment group imbalance over time, which can obscure a true causal effect,\citep{FayLi2024, Hernan2010} or non-proportional hazards. This is particularly relevant for the treatment policy strategy considered here, since non-proportional hazards arise between arms by construction whenever occurrence of the IE modifies the post-IE hazard for $Y$. The average hazard ratio defined in equation~\eqref{eq:DeltaP} is used to approximate the power of the Cox model analysis routinely applied in this setting, but its use does not resolve the underlying issue of non-proportional hazards. Extending our methods to alternative summary measures, such as weighted HRs (e.g., using survival function-based weights as proposed by Schemper et al.\citep{Schemper2009}) or the restricted mean survival time (RMST),\citep{RoystonParmar2013} is an important area for future research.

We also assumed a complete data scenario, where participants either complete the trial or reach the endpoint, without accounting for study withdrawals. In reality participant withdrawal is common, necessitating assumptions about missing data. For composite and hypothetical strategies, where only data up to the IE (or outcome) are considered, study withdrawal might coincide with an IE such as treatment discontinuation, rendering the missing data issue less central. However, for the treatment policy strategy, participants might withdraw after an IE, creating a more complex missing data problem. One approach could be to treat post-IE study withdrawal in a similar manner to an IE with hypothetical handling and use the approach described in Section 2.3.3. This assumes data to be missing-at-random (MAR), which may not be plausible if participants who withdraw are likely to have systematically different outcomes to participants remaining in the trial, which will be a mix of participants with and without an IE. An off-treatment imputation assumption, such as that proposed by Hartley et al.,\citep{Hartley2022} may be a more plausible alternative. Incorporating these missing data assumptions into estimand-specific power calculations requires further investigation.

Finally, the formulae in this paper were initially developed for trials with a fixed follow-up period, such as in the presented nasal polyps case study. Our simulation study showed that the formulae can also approximate power very closely in settings with variable follow-up durations by using the expected follow-up length for $\tau$. Many time-to-event trials, particularly event-driven studies common in oncology, involve variable follow-up until a pre-specified number of events is observed. Extending our methodology to this setting presents challenges, including the need to incorporate participant accrual patterns and the interplay between event rates and trial duration, and is an area of ongoing work.

\section{Conclusion}

Assumptions about IEs and their impact on power should be carefully considered in practice, and sensitivity of sample size requirements to these assumptions should be thoroughly investigated. The formulae proposed in this paper facilitate this, by allowing many scenarios to be investigated for different trial designs much more quickly than simulation-based approaches, which can also be complex to code particularly for estimands with multiple IEs. Finally, our proposed methods extend naturally to assurance calculations, aligning them to the estimand(s) of interest by placing prior distributions on the assumed IE rates and post-IE outcomes, thus incorporating uncertainty in these design parameters into the calculations.

\section*{Acknowledgements} 
The authors would like to thank Adrian Mander for helpful comments on an earlier draft of this manuscript. 

\section*{Conflicts of Interest} 
DJB, SR and TD are employees of and hold shares in GSK. FG holds shares in GSK. 

\bibliographystyle{unsrtnat} 
\bibliography{references} 

\clearpage 
\appendix
\renewcommand{\thefigure}{A\arabic{figure}}
\setcounter{figure}{0}
\renewcommand{\thetable}{A\arabic{table}}
\setcounter{table}{0}

\section*{Supplementary Material} 

\addcontentsline{toc}{section}{Supplementary Material} 

\subsection*{Appendix A: Hypothetical Strategy}

For a single IE with hypothetical handling strategy, the probability of $Y$ being observed prior to time $t$ and before $E$ is 
\begin{align*} 
	P(T^{\mathcal{H}}_{ij}<t, Y^{\mathcal{H}}_{ij}=1) &= \int_0^t P(T^E_{ij}>x) f(x;\lambda_j)\,dx \\ 
													&= \lambda_j\int_0^t e^{-(\lambda_j+\kappa_j)x}\,dx \\ 
													&= \frac{\lambda_j}{\lambda_j+\kappa_j}\left(1-e^{-(\lambda_j+\kappa_j)t}\right) \\
													&= \frac{\lambda_j}{\lambda_j+\kappa_j}F(t;\lambda_j+\kappa_j). 
\end{align*} 

\subsection*{Appendix B: Treatment Policy Strategy} 

Given the memorylessness property of the exponential distribution, and the assumption that the rate parameter for $T_{ij}^Y$ changes from $\lambda_j$ to $\lambda_j^*$ at time $T_{ij}^E=s$, the probability of $Y$ occurring by time $t$ given $E$ occurred at time $s<t$ is
\begin{align*}
P\!\left(T_{ij}^Y \le t \mid T_{ij}^E=s\right)
&= 1 - P\!\left(T_{ij}^Y > t \mid T_{ij}^E=s\right) \\
&= 1 - P\!\left(T_{ij}^Y > s \mid T_{ij}^E=s\right)
      P\!\left(T_{ij}^Y > t \mid T_{ij}^Y > s,\, T_{ij}^E=s\right) \\
&= 1 - \bigl(1 - F(s;\lambda_j)\bigr)\,
      \frac{1 - F(t;\lambda_j^*)}{1 - F(s;\lambda_j^*)} \\
&= 1 - \bigl(1 - F(s;\lambda_j)\bigr)\,\bigl(1 - F(t-s;\lambda_j^*)\bigr) \\
&= 1 - e^{-(\lambda_j s + \lambda_j^*(t-s))}.
\end{align*}

This is equivalent to a piecewise exponential distribution for $T_{ij}^Y \mid T_{ij}^E=s$ where the value of the hazard changes from
$\lambda_j$ to $\lambda_j^*$ at time $T_{ij}^E=s$.

The probability of $Y$ and $E$ occurring prior to $t$ is then:
\begin{align*}
P\!\left(T_{ij}^Y \le t,\, T_{ij}^E \le t\right)
&= \int_{0}^{t} P\!\left(T_{ij}^Y \le t \mid T_{ij}^E=s\right)\, f(s;\kappa_j)\, ds \\
&= \int_{0}^{t} \Bigl(1 - e^{-(\lambda_j s + \lambda_j^*(t-s))}\Bigr)\, \kappa_j e^{-\kappa_j s}\, ds \\
&= \kappa_j \int_{0}^{t} \left(e^{-\kappa_j s} - e^{-(\lambda_j s + \lambda_j^*(t-s) + \kappa_j s)}\right)\, ds \\
&= \kappa_j \left[-\frac{e^{-\kappa_j s}}{\kappa_j}\right]_{0}^{t}
 + \kappa_j \left[\frac{e^{-(\lambda_j s + \lambda_j^*(t-s) + \kappa_j s)}}{\lambda_j - \lambda_j^* + \kappa_j}\right]_{0}^{t} \\
&= 1 - e^{-\kappa_j t} + \frac{\kappa_j}{\lambda_j - \lambda_j^* + \kappa_j}
\left(e^{-(\lambda_j + \kappa_j)t} - e^{-\lambda_j^* t}\right).
\end{align*}

\subsection*{Appendix C: Hypothetical \& Treatment Policy Strategies} 

Here we consider an estimand which applies a hypothetical strategy for $E_1$ and a treatment policy strategy for $E_2$, and derive the probability of observing $Y$ before both time $t$ and $E_1$:
\begin{align*}
F_j^{\mathcal{H},\mathcal{P}}(t)
&= P\!\left(T_{ij}^{\mathcal{H},\mathcal{P}} \le t,\; Y_{ij}^{\mathcal{H},\mathcal{P}}=1\right) \\
&= P\!\left(T_{ij}^Y \le t,\; T_{ij}^Y \le T_{ij}^{E_1},\; T_{ij}^{E_2} > t\right)
 + P\!\left(T_{ij}^Y \le t,\; T_{ij}^Y \le T_{ij}^{E_1},\; T_{ij}^{E_2} \le t\right).
\end{align*}

Using a similar derivation to that in Section 2.2.2 of the main paper for an estimand applying a hypothetical strategy to a single IE, the probability of $Y$ occurring by time $t$ and prior to $E_1$ but with $E_2$ occurring after $t$ is
\begin{align*}
P\!\left(T_{ij}^Y \le t,\; T_{ij}^Y \le T_{ij}^{E_1},\; T_{ij}^{E_2} > t\right)
&= \frac{\lambda_j}{\lambda_j+\kappa_{1j}}\, F(t;\lambda_j+\kappa_{1j})\bigl(1-F(t;\kappa_{2j})\bigr) \\
&= \frac{\lambda_j}{\lambda_j+\kappa_{1j}} \left(1-e^{-(\lambda_j+\kappa_{1j})t}\right)e^{-\kappa_{2j} t}.
\end{align*}

Next, the probability of $Y$ occurring prior to both time $t$ and $E_1$, and $E_2$ occurring prior to time $t$ is
\begin{align*}
P\!\left(T_{ij}^Y \le t,\; T_{ij}^Y \le T_{ij}^{E_1},\; T_{ij}^{E_2} \le t\right)
&= \int_{0}^{t} P\!\left(T_{ij}^Y \le t,\; T_{ij}^Y \le T_{ij}^{E_1} \mid T_{ij}^{E_2}=s\right) f(s;\kappa_{2j})\, ds .
\end{align*}

To derive the probability of $Y$ occurring prior to both $E_1$ and $t$, conditioning on $E_2$ occurring at time $s<t$, first note that, as for a single IE with treatment policy strategy (Appendix B), event times $T_{ij}^Y \mid T_{ij}^{E_2}=s$ follow a piecewise exponential distribution with density function
\begin{align*}
f_{T_{ij}^Y \mid T_{ij}^{E_2}=s}(x)
=
\begin{cases}
\lambda_j e^{-\lambda_j x}, & 0 < x < s, \\
e^{-\lambda_j s}\,\lambda_j^* e^{-\lambda_j^*(x-s)}, & x > s.
\end{cases}
\end{align*}

Also noting that $P(T_{ij}^{E_1} > x)=e^{-\kappa_{1j} x}$ if $x<s$ and $P(T_{ij}^{E_1} > x)=e^{-\kappa_{1j}^* x}$ otherwise, and letting $\nu_j = \lambda_j + \kappa_{1j}$ and $\nu_j^* = \lambda_j^* + \kappa_{1j}^*$ be the hazard for $\min(T_{ij}^Y, T_{ij}^{E_1})$, pre and post occurrence of $E_2$ respectively, the probability of $Y$ occurring prior to both $E_1$ and $t$, conditioning on $E_2$ occurring at time $s<t$ is
\begin{align*}
P\!\left(T_{ij}^Y \le t,\; T_{ij}^Y \le T_{ij}^{E_1} \mid T_{ij}^{E_2}=s\right)
&= \int_{0}^{t} P\!\left(T_{ij}^{E_1} > x\right)\, f_{T_{ij}^Y \mid T_{ij}^{E_2}=s}(x)\, dx \\
&= \int_{0}^{s} \lambda_j e^{-(\lambda_j+\kappa_{1j})x}\, dx
 + \int_{s}^{t} \left[e^{-\lambda_j s}\lambda_j^* e^{-\lambda_j^*(x-s)} e^{-\kappa_{1j}^* x}\right] dx \\
&= \frac{\lambda_j}{\nu_j}\left(1-e^{-\nu_j s}\right)
 + e^{-\nu_j s}\frac{\lambda_j^*}{\nu_j^*}\left(1-e^{-\nu_j^*(t-s)}\right).
\end{align*}

We can then calculate
\begin{align*}
P\!\left(T_{ij}^Y \le t,\; T_{ij}^Y \le T_{ij}^{E_1},\; T_{ij}^{E_2} \le t\right)
&= \int_{0}^{t} P\!\left(T_{ij}^Y \le t,\; T_{ij}^Y \le T_{ij}^{E_1} \mid T_{ij}^{E_2}=s\right) f(s;\kappa_{2j})\, ds
\end{align*}
by splitting the integral into two parts. The first term is
\begin{align*}
I_1
&= \frac{\lambda_j}{\nu_j} \int_{0}^{t} \left(1-e^{-\nu_j s}\right)\kappa_{2j} e^{-\kappa_{2j}s}\, ds \\
&= \frac{\lambda_j}{\nu_j}
\left[
\int_{0}^{t} \kappa_{2j} e^{-\kappa_{2j}s}\, ds
-
\int_{0}^{t} \kappa_{2j} e^{-(\nu_j+\kappa_{2j})s}\, ds
\right] \\
&= \frac{\lambda_j}{\nu_j}
\left[
\left(1-e^{-\kappa_{2j}t}\right)
-\frac{\kappa_{2j}}{\nu_j+\kappa_{2j}}\left(1-e^{-(\nu_j+\kappa_{2j})t}\right)
\right].
\end{align*}

The second term is
\begin{align*}
I_2
&= \frac{\lambda_j^*}{\nu_j^*}\int_{0}^{t} e^{-\nu_j s}\left(1-e^{-\nu_j^*(t-s)}\right)\kappa_{2j}e^{-\kappa_{2j}s}\, ds \\
&= \frac{\lambda_j^*}{\nu_j^*}\kappa_{2j}
\left[
\int_{0}^{t} e^{-(\nu_j+\kappa_{2j})s}\, ds
-
\int_{0}^{t} e^{-(\nu_j+\kappa_{2j})s}e^{-\nu_j^*(t-s)}\, ds
\right].
\end{align*}

The first integral in $I_2$ is
\begin{align*}
\int_{0}^{t} e^{-(\nu_j+\kappa_{2j})s}\, ds
= \frac{1-e^{-(\nu_j+\kappa_{2j})t}}{\nu_j+\kappa_{2j}}.
\end{align*}
The second integral is
\begin{align*}
\int_{0}^{t} e^{-(\nu_j+\kappa_{2j})s}e^{-\nu_j^*(t-s)}\, ds
&= e^{-\nu_j^* t}\int_{0}^{t} e^{-[(\nu_j+\kappa_{2j})-\nu_j^*]s}\, ds \\
&= e^{-\nu_j^* t}\,\frac{1-e^{-(\nu_j-\nu_j^*+\kappa_{2j})t}}{\nu_j-\nu_j^*+\kappa_{2j}}.
\end{align*}

Therefore,
\begin{align*}
P\!\left(T_{ij}^Y \le t,\; T_{ij}^Y \le T_{ij}^{E_1},\; T_{ij}^{E_2} \le t\right)
&= I_1 + I_2 \\
&= \frac{\lambda_j}{\nu_j}
\left[
\left(1-e^{-\kappa_{2j}t}\right)
-\frac{\kappa_{2j}}{\nu_j+\kappa_{2j}}\left(1-e^{-(\nu_j+\kappa_{2j})t}\right)
\right] \nonumber\\
&\quad + \frac{\lambda_j^*}{\nu_j^*}\kappa_{2j}
\left[
\frac{1-e^{-(\nu_j+\kappa_{2j})t}}{\nu_j+\kappa_{2j}}
- e^{-\nu_j^* t}\frac{1-e^{-(\nu_j-\nu_j^*+\kappa_{2j})t}}{\nu_j-\nu_j^*+\kappa_{2j}}
\right].
\end{align*}

Finally, the probability of observing $Y$ before both time $t$ and $E_1$ is
\begin{align*}
F_j^{\mathcal{H},\mathcal{P}}(t)
&= P\!\left(T_{ij}^{\mathcal{H},\mathcal{P}} \le t,\; Y_{ij}^{\mathcal{H},\mathcal{P}}=1\right) \\
&= P\!\left(T_{ij}^Y \le t,\; T_{ij}^Y \le T_{ij}^{E_1},\; T_{ij}^{E_2} > t\right)
 + P\!\left(T_{ij}^Y \le t,\; T_{ij}^Y \le T_{ij}^{E_1},\; T_{ij}^{E_2} \le t\right) \\
&= \frac{\lambda_j}{\nu_j}\left(1-e^{-\nu_j t}\right)e^{-\kappa_{2j}t}
+ \frac{\lambda_j}{\nu_j}
\left[
\left(1-e^{-\kappa_{2j}t}\right)
-\frac{\kappa_{2j}}{\nu_j+\kappa_{2j}}\left(1-e^{-(\nu_j+\kappa_{2j})t}\right)
\right] \nonumber\\
&\quad + \frac{\lambda_j^*}{\nu_j^*}\kappa_{2j}
\left[
\frac{1-e^{-(\nu_j+\kappa_{2j})t}}{\nu_j+\kappa_{2j}}
- e^{-\nu_j^* t}\frac{1-e^{-(\nu_j-\nu_j^*+\kappa_{2j})t}}{\nu_j-\nu_j^*+\kappa_{2j}}
\right] \\
&= \frac{\lambda_j}{\nu_j+\kappa_{2j}}\left(1-e^{-(\nu_j+\kappa_{2j})t}\right)
+ \frac{\lambda_j^*}{\nu_j^*}\kappa_{2j}
\left[
\frac{1-e^{-(\nu_j+\kappa_{2j})t}}{\nu_j+\kappa_{2j}}
- e^{-\nu_{1j}^* t}\frac{1-e^{-(\nu_j-\nu_j^*+\kappa_{2j})t}}{\nu_j-\nu_j^*+\kappa_{2j}}
\right].
\end{align*}

\subsection*{Appendix D: Three IEs}

Suppose an estimand has three IEs $E_1$, $E_2$ and $E_3$ in which a composite strategy is used for $E_1$, a hypothetical strategy is used for $E_2$ and a treatment policy strategy is used for $E_3$ ($\mathcal{S} = \mathcal{C}, \mathcal{H}, \mathcal{P}$). As before, we assume that occurrence of $E_3$ may modify the risk of $Y$ and the other IEs, $E_1$ and $E_2$. Assuming that event times for each IE are independent of each other and those for $Y$ (conditional on whether $E_3$ has occurred or not), then $E_2$ again creates an independent censoring process for the composite endpoint of $Y$ and $E_1$ and so $\Delta^{\mathcal{C},\mathcal{H},\mathcal{P}} = \Delta^{\mathcal{C},\mathcal{P}}$.

The event time and endpoint indicator is defined in a similar manner as that for the combined composite and hypothetical strategies, i.e.\ $T_{ij}^{\mathcal{C},\mathcal{H},\mathcal{P}} = \min(T_{ij}^Y, T_{ij}^{E_1}, T_{ij}^{E_2})$ and $Y_{ij}^{\mathcal{C},\mathcal{H},\mathcal{P}} = \mathbb{I}(\min(T_{ij}^Y, T_{ij}^{E_1}) \leq T_{ij}^{E_2})$. The probability of observing the event $Y_{ij}^{\mathcal{C},\mathcal{H},\mathcal{P}}$ by time $t$ can be derived in a similar manner to the analogous probability in Section 2.3.3 of the main paper for the hypothetical and treatment policy strategies, but replacing the hazard for $Y$ by the corresponding hazard for the composite endpoint ($Y$ or $E_1$) so that
\[
F_j^{\mathcal{C},\mathcal{H},\mathcal{P}}(t) = \frac{\lambda_j + \kappa_{1j}}{\nu_j + \kappa_{3j}}\!\left(1 - e^{-(\nu_j + \kappa_{3j})t}\right)
+ \frac{\lambda_j^* + \kappa_{1j}^*}{\nu_j^*}\kappa_{3j}\!\left[\frac{1 - e^{-(\nu_j + \kappa_{3j})t}}{\nu_j + \kappa_{3j}} - e^{-\nu_j^* t}\frac{1 - e^{-(\nu_j - \nu_j^* + \kappa_{3j})t}}{\nu_j - \nu_j^* + \kappa_{3j}}\right]
\]
where $\nu_j = \lambda_j + \kappa_{1j} + \kappa_{2j}$ is now the combined hazard for $Y$, $E_1$ and $E_2$, and event hazards post $E_3$ are denoted with an asterisk. The expected number of composite events ($Y$ or $E_1$) observed on arm $j$ by time $\tau$ is then $D_j^{\mathcal{C},\mathcal{H},\mathcal{P}} = n_j F_j^{\mathcal{C},\mathcal{H},\mathcal{P}}(\tau)$.

\subsection*{Appendix E}

\textbf{Simulation results: one IE, with IE times independent of pre-IE event times for outcome $Y$ (exponentially distributed event times)}

\begin{figure}[H] \centering
	\includegraphics[width=0.95\textwidth]{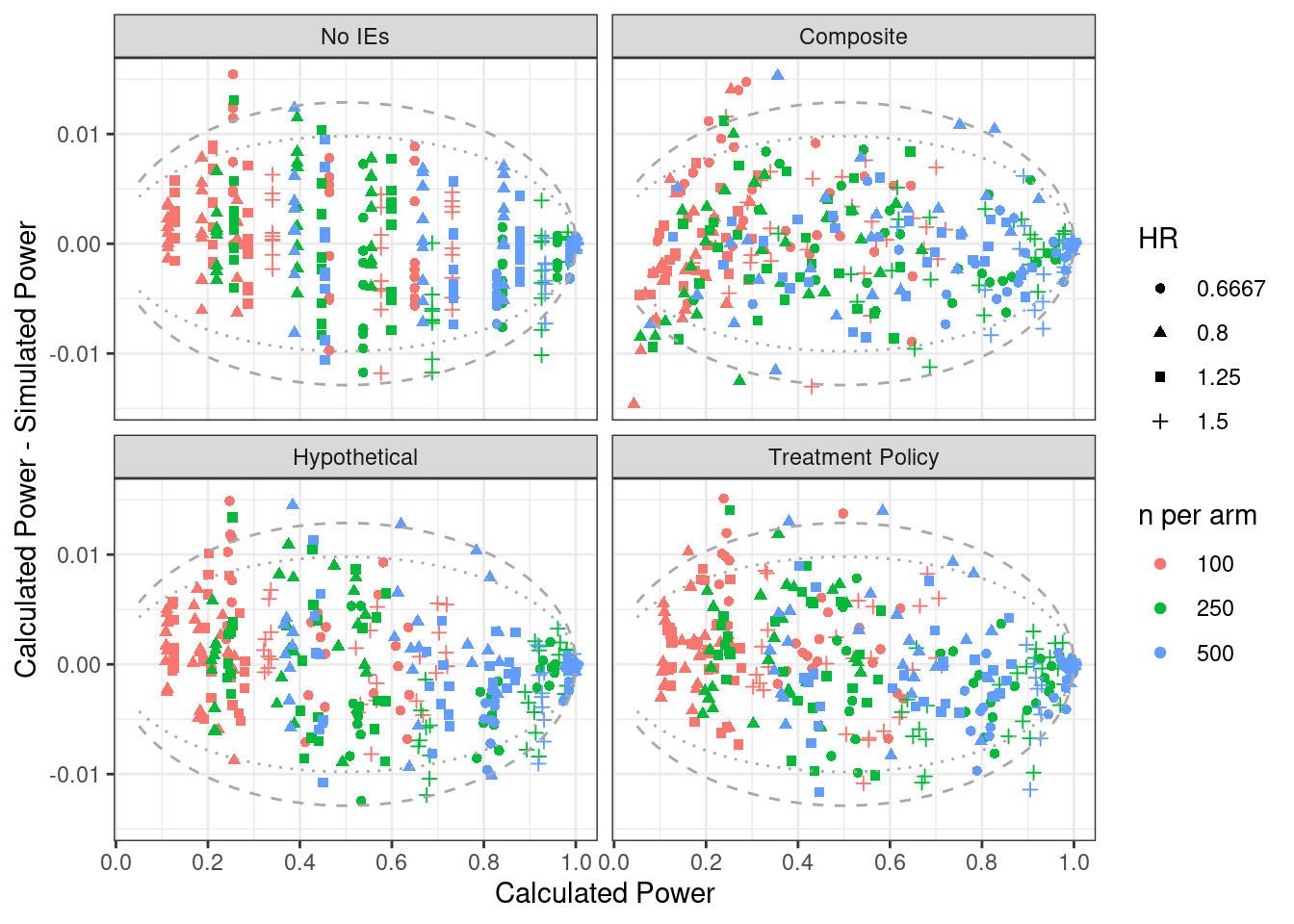}
	\caption{Difference between calculated and simulated power versus calculated power for each design scenario and IE strategy, for estimands with one IE. Short-dashed and long-dashed lines show the expected 95\% and 99\% CIs respectively for the power difference, assuming no true difference in power between simulated and calculated values.} \label{fig:A1}
\end{figure}

\clearpage
\subsection*{Appendix F}

\textbf{Simulation study: Two IEs}

A simulation study was conducted for estimands with two IEs, $E_1$ and $E_2$, assumed to be independent each other and of outcome $Y$, and handled with the following combination of strategies:
\begin{enumerate}
  \item[(a)] Composite ($E_1$) and hypothetical ($E_2$)
  \item[(b)] Composite ($E_1$) and treatment policy ($E_2$)
  \item[(c)] Hypothetical ($E_1$) and treatment policy ($E_2$)
\end{enumerate}

The same design parameters used for $n$, $\lambda_0$ and $\Delta^Y$ in the simulation study with one IE were used. Event times for $Y$, $E_1$ and $E_2$ were assumed to be exponentially distributed. In addition, by setting the IE rate for IE $E_l$ at a proportion $\pi_l$ of the corresponding event rate for the endpoint on the control arm, the following hazards for each IE were considered:
\begin{itemize}
  \item Hazard for $E_1$: $\kappa_{1j} = -\log(1 - q_{10})/\tau$, where $q_{10} = \pi_1 p_0$ and $\pi_1 \in \{0.1, 0.25\}$ $(j=0,1)$
  \item Hazard for $E_2$: $\kappa_{2j} = -\log(1 - q_{20})/\tau$, where $q_{20} = \pi_2 p_0$ and $\pi_2 \in \{0.1, 0.25\}$ $(j=0,1)$
\end{itemize}

In total 144 design scenarios were assessed for each combination of IE strategies. Event times for the two IEs, $T_{ij}^{E_1}$ and $T_{ij}^{E_2}$, were drawn from independent exponential distributions with rate parameters $\kappa_{1j}$ and $\kappa_{2j}$ respectively for treatment $j$ ($j=0,1$).

In scenarios (b) and (c) where a treatment policy strategy is used for $E_2$, the hazards for both $Y$ and $E_1$ on the active arm are assumed to switch to the corresponding hazards on placebo following occurrence of $E_2$. Post-$E_2$ event times for $Y$ and $E_1$, denoted $T_{i1}^{Y^*}$ and $T_{i1}^{E_1^*}$ respectively, were therefore drawn from independent exponential distributions with parameters $\lambda_0$ and $\kappa_{10}$ respectively for participants on treatment $j=1$ for whom the event time on $E_2$ occurred prior to event times originally simulated for $Y$ and $E_1$, i.e.\ $T_{i1}^{E_2} < \min(T_{i1}^Y, T_{i1}^{E_1})$. The event times and event indicators used for each combination of strategies are summarised in Table~\ref{tab:A1}.

\begin{figure}[H] \centering
	\includegraphics[width=0.95\textwidth]{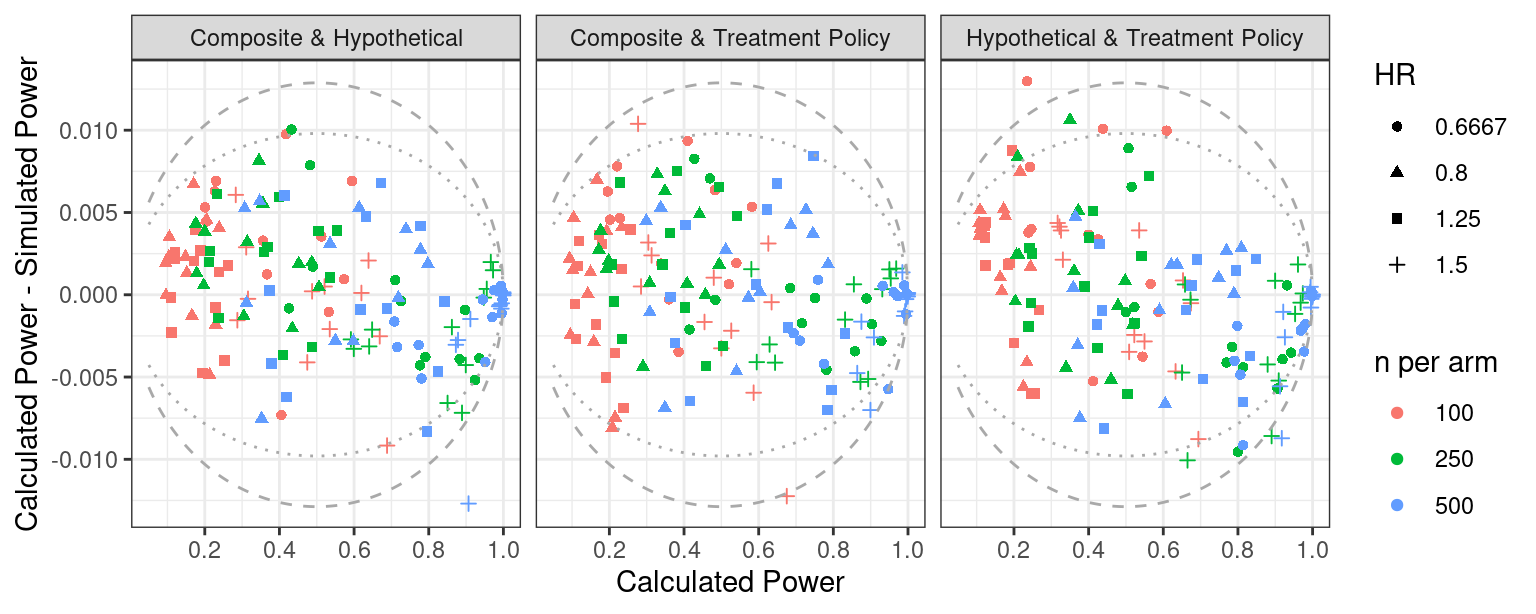}
	\caption{Difference between calculated and simulated power versus calculated power for each design scenario and IE strategy combination, for estimands with two IEs. Short-dashed and long-dashed lines show the expected 95\% and 99\% CIs respectively for the power difference, assuming no true difference in power between simulated and calculated values.} \label{fig:A2}
\end{figure}

Figure~\ref{fig:A2} shows the difference between calculated and simulated power against calculated power for all design scenarios and combination of IE strategies assessed. As with the simulations of estimands with a single IE, estimates were generally in agreement up to MC error and there were no systematic differences identified across any of the design parameters considered.

\begin{sidewaystable}[p]
\centering
\caption{Definition of event times and indicators for each combination of IE strategies assessed in simulations of estimands with two IEs.}
\label{tab:A1}
\small
\begin{tabular}{p{2.4cm} p{1.2cm} p{7cm} p{7cm}}
\toprule
\textbf{Strategy Combination} $(\mathcal{S})$ \textbf{for 2 IEs} & & \textbf{Event Time,} $T_{ij}^{\mathcal{S}}$ & \textbf{Event Indicator by time} $\tau$, $Y_{ij}^{\mathcal{S}}$ \\
\midrule
$\mathcal{C},\mathcal{H}$ & & $\min(T_{ij}^{E_1}, T_{ij}^{E_2}, T_{ij}^Y, \tau)$ & $\mathbb{I}\!\left(\min(T_{ij}^{E_1}, T_{ij}^Y) < \min(T_{ij}^{E_2}, \tau)\right)$ \\[6pt]
\multirow{4}{*}{$\mathcal{C},\mathcal{P}$} & $j=0$: & $\min(T_{i0}^{E_1}, T_{i0}^Y, \tau)$ & $\mathbb{I}(T_{ij}^{\mathcal{C},\mathcal{P}} < \tau)$ \\
 & $j=1$: & $\min(T_{i1}^Y, T_{i1}^{E_1}, \tau)$, if $\min(T_{i1}^Y, T_{i1}^{E_1}) < T_{i1}^{E_2}$ & \\
 & & $\min(T_{i1}^{E_2}+T_{i1}^{Y^*}, T_{i1}^{E_2}+T_{i1}^{E_1^*}, \tau)$, otherwise & \\[6pt]
\multirow{5}{*}{$\mathcal{H},\mathcal{P}$} & $j=0$: & $\min(T_{i0}^{E_1}, T_{i0}^Y, \tau)$ & $\mathbb{I}(T_{i0}^Y < \min(T_{i0}^{E_1}, \tau))$ \\
 & $j=1$: & $\min(T_{i1}^Y, T_{i1}^{E_1}, \tau)$, if $\min(T_{i1}^Y, T_{i1}^{E_1}) < T_{i1}^{E_2}$ & $\mathbb{I}(T_{ij}^Y < \min(T_{ij}^{E_1}, \tau))$, if $\min(T_{i1}^Y, T_{i1}^{E_1}) < T_{i1}^{E_2}$ \\
 & & $\min(T_{i1}^{E_2}+T_{i1}^{Y^*}, T_{i1}^{E_2}+T_{i1}^{E_1^*}, \tau)$, otherwise & $\mathbb{I}(T_{i1}^{E_2}+T_{i1}^{Y^*} < \min(T_{i1}^{E_2}+T_{i1}^{E_1^*}, \tau))$, otherwise \\
\bottomrule
\end{tabular}
\end{sidewaystable}

 \end{document}